\documentclass[prd,preprint,tightenlines,floatfix,preprintnumbers,showpacs,nofootinbib,eqsecnum,draft]{revtex4}
 \usepackage{color}

 \usepackage[dvips,final]{graphicx}
  \usepackage{amssymb,amsmath,bm,pifont}

\newcommand{\gev}[1]{\relax\ifmmode{\text{GeV}^{#1}}              
                     \else{GeV$^{#1}${ }}\fi}                     
\newcommand{\Ds}{\displaystyle}

\newcommand{\nn}{\nonumber}
\def\muF{\relax\ifmmode\mu_\text{F}^2\else{$\mu_\text{F}^2${ }}\fi}
\def\muR{\relax\ifmmode\mu_\text{R}^2\else{$\mu_\text{R}^2${ }}\fi}
\def\muO{\relax\ifmmode{\mu_{0}^{2}}\else{$\mu_{0}^{2}${ }}\fi}

\begin{document}
\thispagestyle{empty}
\date{\today}
\preprint{\hbox{RUB-TPII-02/2010}}

\vspace*{-10mm}
\title{Endpoint behavior of the pion distribution amplitude in QCD
       sum rules with nonlocal condensates}
\author{S. V. Mikhailov}
 \email{mikhs@theor.jinr.ru}
\author{A.~V.~Pimikov}
 \email{pimikov@theor.jinr.ru}
 \affiliation{Bogoliubov Laboratory of Theoretical Physics,
              JINR, 141980 Dubna, Moscow region, Russia\\ }
\author{N.~G.~Stefanis}
 \email{stefanis@tp2.ruhr-uni-bochum.de}
 \affiliation{Institut f\"{u}r Theoretische Physik II,
              Ruhr-Universit\"{a}t Bochum, D-44780 Bochum, Germany\\}
\vspace {10mm}
\begin{abstract}
Starting from the QCD sum rules with nonlocal condensates for the
pion distribution amplitude, we derive another sum rule for its
derivative and its ``integral'' derivatives---defined in this work.
We use this new sum rule to analyze the fine details of the pion
distribution amplitude in the endpoint region $x\sim 0$.
The results for endpoint-suppressed and flat-top (or flat-like)
pion distribution amplitudes are compared with those we obtained
with differential sum rules by employing two different models for
the distribution of vacuum-quark virtualities.
We determine the range of values of the derivatives of the
pion distribution amplitude and show that endpoint-suppressed
distribution amplitudes lie within this range, while those with
endpoint enhancement---flat-type or CZ-like---yield values outside
this range.

\end{abstract}

\maketitle

\section{Introduction}
\label{sec:intro}

Many of the theoretical and phenomenological analyses of QCD processes
rely upon the factorization of the underlying dynamics into a
short-distance dominated part, amenable to QCD perturbation theory,
and a large-distance part that has to be taken from experiment or be
determined by nonperturbative methods.
Among such processes, the pion form factors (electromagnetic and
transition) play the role of a theoretical laboratory to test various
ideas and techniques.
The key ingredient in these descriptions is the pion distribution
amplitude (DA) $\varphi_{\pi}(x)$ which represents the pion bound state.
At leading twist two it is defined in terms of a nonlocal axial
current and reads \cite{Rad77}
\begin{equation}
  \langle 0 |\bar{d}(z) \gamma^{\mu}\gamma_{5} [z,0] u(0)
            | \pi (P)
  \rangle |_{z^{2}=0}
=
   i f_{\pi}P^{\mu} \int_{0}^{1} dx\, {\rm e}^{ix(z\cdot P)}
   \varphi_{\pi}^{\rm (t=2)}(x,\mu_{0}^{2}),
\label{eq:pi-DA}
\end{equation}
where \hbox{$x$ ($\bar{x}\equiv 1-x$)} is the longitudinal momentum
fraction carried by the valence quark (antiquark) in the pion and the
path-ordered exponential, i.e., the lightlike gauge link
\begin{equation}
  [z,0]
=
  {\cal P} \exp \left[ -ig \int_{0}^{z} dy^{\mu} t^{a}A_{\mu}^{a}(y)
                \right] \, ,
\label{eq:con}
\end{equation}
ensures gauge invariance.

The pion DA has an expansion in the basis of the Gegenbauer
polynomials which constitute the eigenfunctions of the one-loop meson
evolution equation.
At a typical hadronic scale $\mu_0^2$, which serves as a normalization
scale, one then has
\cite{LB79}
\begin{equation}
   \varphi^{\rm (t=2)}(x; \mu_0^2)
=
   \varphi^{\rm as}(x)
                      \left[1 + a_2(\mu_0^2)\ C^{3/2}_2(2x-1)
   + a_4(\mu_0^2)\ C^{3/2}_4(2x-1)
       \right] + \ldots \ ,
\label{eq:pion-DA}
\end{equation}
where $\varphi^{\rm as}(x)=6x\bar{x}$ is the asymptotic pion DA.
By virtue of the leptonic decay
$\pi\to\mu^{+}\nu_{\mu}$, one obtains the normalization
$\int_{0}^{1} dx\ \varphi_{\pi}^{\rm (t=2)}(x,\mu_{0}^{2})=1$,
which fixes $a_0=1$.

Rather than try to derive the pion DA as a whole, one attempts to
reconstruct it from its first few moments
\begin{equation}
  \langle \xi^{N} \rangle_{\pi}
\equiv
  \int_{0}^{1} dx (2x-1)^{N} \varphi_{\pi}(x)\, ,
\label{eq:moments}
\end{equation}
where $\xi\equiv 2x-1$.
The values of the moments may be determined by means of QCD sum rules
(SR)s with local \cite{CZ84} or nonlocal condensates
\cite{MR86,MR90,BR91,MR92}, or be computed by numerical
simulations on the lattice \cite{DelDebbio05,Lat06,Lat07}.
Once they are known, one can use them to reverse engineer the pion DA,
with a precision depending upon the influence of the magnitude of the
discarded higher-order moments.
It was shown in Ref.\ \cite{BMS01} that, using QCD sum rules
with nonlocal condensates, one can de facto resort to the first
two Gegenbauer coefficients $a_2, a_4$, while $a_i$ with $i=6, 8, 10$
turn out to be negligible.
Once the shape of the pion DA has been determined at some (low)
normalization scale $\mu^2$ around 1~GeV$^2$, one can evolve the
Gegenbauer coefficients to higher values of the momentum scale using
the Efremov-Radyushkin-Brodsky-Lepage \cite{LB79}
evolution equation which is determined by means of QCD perturbation
theory.

It turns out that another quantity which is intertwined with the
form factors of the pion is its inverse moment
\begin{equation}
  \langle x^{-1} \rangle_{\pi}
=
  \int_{0}^{1} dx \ \frac{1}{x} \ \varphi_{\pi}(x) \ .
\label{eq:inv-mom}
\end{equation}
This quantity is one of the key ingredients of the pion-photon
transition form factor, a process that has attracted the continuous
attention of theorists
\cite{KMR86,KR96,RR96,MuR97,SSK99,SY99,SSK00,BMS01,MNP01a,%
DKV01,MMP02,BMS02,BMS03,Ag05b,BMS05lat,Ste08,MS09}
and experimentalists \cite{CELLO91,CLEO98,BaBar09}.
Actually, the most recent measurement of this observable by the BaBar
Collaboration \cite{BaBar09} has provided controversial results.
At moderate values of the momentum transfer, up to 10~GeV$^2$, the new
high-precision BaBar data agree well with the previous CLEO data
\cite{CLEO98} and can be best described by pion DAs that have their
endpoints strongly suppressed \cite{MS09,MS09Trento,MS09strba}---as the
Bakulev-Mikhailov-Stefanis (BMS) model \cite{BMS01} derived from QCD
sum rules with nonlocal condensates.
By contrast, the high-$Q^2$ BaBar data show an unexpected growth with
$Q^2$ which cannot be understood on the basis of collinear
factorization and calls for pion DAs that have their endpoints strongly
enhanced \cite{Rad09,Pol09}.
This intriguing behavior consists the basic motivation for the present
investigation, though we will not attempt to describe any data.

We shall employ in this work the method of QCD SRs with nonlocal
condensates (NLC)s with the aim to estimate the slope of the
pion DA in the region $x \sim 0$, trying to understand the fine
structure of the pion DA in this region vs. the ansatz for the
quark-virtuality distribution in the nonperturbative QCD vacuum.
Our main interest will be in the behavior of pion DAs with distinct
endpoint characteristics.
QCD SRs were mainly proposed with the purpose of studying the
\emph{integral} characteristics of the pion DA.
To overcome this restriction, we shall design an operator for defining
\emph{integral derivatives} of the pion DA.
These will supplement the results obtained with SRs which employ the
\emph{standard derivative} of the pion DA.
In this latter case, we will use in our analysis not only a
delta-function ansatz for the vacuum quark-virtuality distribution,
but also a refined model which describes the large-distance regime more
accurately.

The paper is organized as follows.
In Sec.\ \ref{sec:QCD-vac}, we briefly discuss the main features of
the nonperturbative QCD vacuum in terms of the scalar quark condensate.
We also give the corresponding SRs and recall their main ingredients.
Section \ref{sec:pi-slope} is devoted to the calculation of the slope
of the pion DA in the endpoint region employing two different
techniques: integral SRs and differential SRs.
Finally, Sec.\ \ref{sec:concl} contains our conclusions, while some
important technical details are given in four appendices.

\section{Nonperturbative QCD vacuum with nonlocal condensates}
\label{sec:QCD-vac}

The basic idea underlying the NLC approach is that the vacuum
condensates possess a correlation length which endows the vacuum quarks
with a non-zero average virtuality $\langle k^2_q\rangle$
(see, for instance, \cite{Rad97}).
To analyze the nonlocality of the vacuum condensate, it is useful
to parameterize the lowest one\footnote{%
In this work we use the gauge $z^\mu A_\mu = 0$.
Therefore, one has $[0,z]=1$.
}
$\langle{\bar{q}(0)[0,z]q(z)}\rangle\equiv M_S(z^2)$
with the help of the vacuum distribution function $f_S(\alpha)$:
\begin{eqnarray}
  M_S(z^2)
&=&
  \langle{\bar{q}q}\rangle
  \int\limits_0^\infty\!\! f_S(\alpha)\,e^{\alpha z^2/4}\,d \alpha
\label{eq:nonlocCon}
\end{eqnarray}
that describes the distribution of the vacuum-quark virtuality
$\alpha$ \cite{MR86}.
Assuming
$
 f_S(\alpha)
=
 \delta(\alpha-\lambda_q^2/2)
$,
that takes into account only a fixed virtuality $\lambda_q^2$ of
the vacuum quarks, leads to the simplest Gaussian model
\begin{eqnarray}
\langle{\bar{q}(0)q(z)}\rangle
=
  \langle\bar{q}(0)q(0)\rangle {\rm e}^{-|z^{2}|\lambda_{q}^{2}/8}
\label{eq:scalar-cond}
\end{eqnarray}
for the scalar quark condensate \cite{MR86}.

The parameter $\lambda_{q}^{2}$ represents the typical quark momentum
in the vacuum and is given by
\begin{eqnarray}
  \langle k_q^2\rangle
=
  \frac{\langle\bar{q}(0)\nabla^{2}q(0)\rangle}
       {\langle\bar{q}(0)q(0)\rangle}
\equiv
  \lambda_{q}^{2} \ .
\label{eq:lambda}
\end{eqnarray}
In this work, we use the value
$\lambda_{q}^{2}=0.4$~GeV$^2$, which is supported by several analyses,
though values within the interval $[0.35\div 0.45]$~GeV$^2$ are still
acceptable (see \cite{BMS02,BMS01,BM02} and references cited therein).

The QCD SRs with nonlocal condensates for the pion DA were first
proposed in \cite{MR86} and were significantly improved in
\cite{BMS01} from which we quote
\begin{eqnarray}
  f_{\pi}^2\,\varphi_\pi(x)
  + f_{A_1}^2\,\varphi_{A_1}\!(x)\, e^{-m_{A_1}^2/M^2}
  + \int\limits_{s_0}^{\infty}
                              \rho_\text{pert}\left(x\right)
                              e^{-s/M^2}ds
=
   \int\limits_{0}^{\infty}
                           \rho_\text{pert}\left(x\right)
                           e^{-s/M^2}ds \nn\\
   + \Delta\Phi_\text{G}(x,M^2)
   + \left[\Delta\Phi_\text{S}(x,M^2)
   + \Delta\Phi_\text{V}(x,M^2)
   + \Delta\Phi_\text{T}(x,M^2)\right]_{\rm Q} \ .
\label{eq:pionDAQCDSR}
\end{eqnarray}
Here, $\varphi_{A_1}$ is the $A_1$-meson DA, whereas
$f_\pi$ and $f_{A_1}$ are, respectively, the decay constants of the
$A_1$ and the $\pi$-meson.
The $A_1$-meson state is an effective state that collects the $\pi'$
and the $a_1$ meson.
The nonperturbative ingredients in the theoretical part of the SR
are the gluon-condensate term
$\Delta\Phi_{G}(x,M^2)$ and the quark-condensate contribution
$\left[...\right]_{\rm Q}$.
This latter contribution contains the vector-condensate term (V),
the mixed quark-gluon condensate term (T), and the scalar condensate
term (S).
The explicit expressions for the nonperturbative contributions and
the NLO spectral density
$\rho_\text{pert}^\text{(NLO)}\left(x\right)$
are given in Appendices \ref{App:A.1} and \ref{App:A.2}, respectively.
It turns out that in the endpoint region, the first radiative
correction in the spectral density, which is of $\mathcal{O}(\alpha_s)$,
is too large, thus overshadowing the zeroth order perturbative
contribution and the nonperturbative contribution.
For that reason, we use in this work the leading-order (LO)
approximation
$\rho_\text{pert}^\text{(LO)}\!\left(x\right)=3x\bar x/2\pi^2$.
In order to include radiative corrections into the spectral
density---when analyzing the endpoint region---one would have to resum
all radiative corrections, a formidable task outside the scope of the
present investigation.

\section{Slope of the pion DA}
\label{sec:pi-slope}

\begin{figure}[h!]
\centerline{\includegraphics[width=0.45\textwidth]{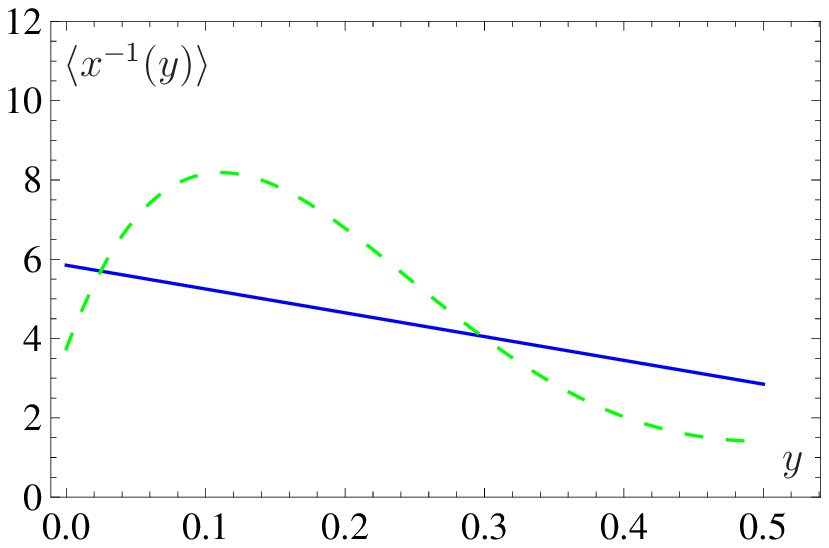}   
         ~~~\includegraphics[width=0.45\textwidth]{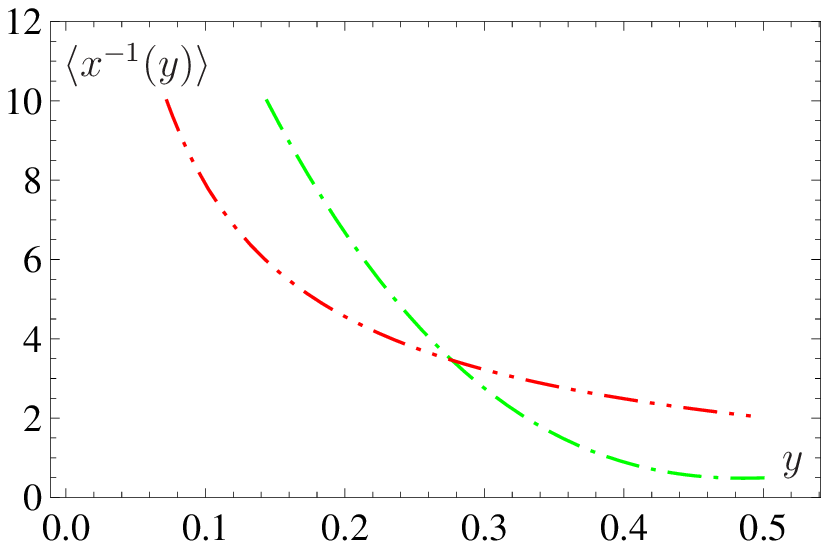}   
         }
\caption{\footnotesize
   Graphical representation of $\langle x^{-1}(y)\rangle$, integrated
   in the region $y$ to $y+\delta$ for different pion DAs.
   Left panel: Solid blue line---\underline{asy}mptotic;
                           dashed green line---BMS \cite{BMS01}.
   Right panel: Dashed-dotted green line---CZ \cite{CZ84};
                           dashed-dotted-dotted red line---flat-top DA
                           [Eq.\ (\ref{eq:flat-DA}) with $\alpha=0.1$].
   \label{fig:inv.mom.y}}
\end{figure}

As already mentioned in the Introduction, the endpoint region of the
pion DA turns out to be of particular importance for a variety of pion
observables---see \cite{SSK99,BPSS04,BMS04kg} for an in-depth
discussion of this issue.
For example, in the case of the pion-photon transition form factor, one
finds that in order to comply with the CLEO data---and those BaBar data
close to them---one needs endpoint-suppressed pion DAs, while the
high-$Q^2$ BaBar data can only be described with flat-type pion DAs.
Therefore, it is crucial to explore the fine details of the pion DA in
a region around the origin $x \sim 0$.
One way to study the endpoint regime of the pion DA is provided by the
inverse moment $\langle x^{-1} \rangle_{\pi}$ \cite{BMS03dur}.
In Fig.\ \ref{fig:inv.mom.y} we show the values of
\begin{eqnarray}
  \langle x^{-1}(y)\rangle
  &=&
  \frac{1}{\delta}\int_y^{y+\delta }\frac{\phi _{\pi }(x)}{x}\,dx
\label{eq:rel-weight}
\end{eqnarray}
with $\delta=0.05$ for some pion DA models with a characteristic
behavior in the endpoint region.
From this figure one sees that the BMS pion DA (dashed green line),
derived with the aid of nonlocal condensates \cite{BMS01},
exhibits an evident endpoint suppression, while all other models have
more (flat-top---dashed-dotted-dotted red line---and CZ---dashed-dotted
green line) or less (asymptotic DA) endpoint enhancement.
The considered models of the pion DA, and their parameters, are given
in Appendix \ref{App:DAmodels}.
In the present work we will probe the endpoint region of the
pion DA also by another means, namely, the ``integral''
derivative which will be defined next.

\subsection{``Integral'' sum rules}
\label{subsec:int-SR}

We introduce now the integral derivatives in order to discuss the slope
of the pion DA.
We define these quantities in terms of a set of operators $D^{(n)}$
given by
\begin{eqnarray}
  [D^{(0)}\varphi](x)
=
  \varphi'(x)\,,~~~
  [D^{(1)}\varphi](x)
=
  \varphi(x)/x\,,~~~
  [D^{(2)}\varphi](x)
=
  \frac{1}{x}\int\limits_0^x\frac{\varphi(y)}{y} dy \ .
\label{eq:int-dev012}
\end{eqnarray}
Then, assuming $\varphi(0)=0$, we get the iterative formula
\begin{eqnarray}
  [D^{(n+1)}\varphi](x)
=
  \frac{1}{x}\int\limits_0^x\!\! dy\,[D^{(n)}\varphi](y) \ .
\label{eq:int-dev-it}
\end{eqnarray}
Thus, each higher derivative within the set of the differential
operators $D^{(n)}$, is stronger averaged with respect to $x$
than the previous one.
The usefulness of these derivatives derives from the fact that they
can be applied to QCD sum rules which in general contain on their
RHS singular contributions.
In Appendix \ref{App:A.3}, we elaborate on $D^{(n)}$
so that here we can focus our attention on the main properties of
these derivatives.
First, it is obvious that $D^{(n)}$ acts on a linear function as a
differentiation operator, i.e.,
$D^{(n)}a x=a$.
Second, assuming that the Taylor expansion of $\varphi(x)$ at $x=0$
exists, one finds from (\ref{eq:D.nu.x})
\begin{eqnarray}
  [D^{(\nu+2)}\varphi](x)
  =
  \varphi'(0)+\varphi''(0)\frac{x}{2!2^{\nu+1}}
  + O\left(\frac{x^2}{3^{\nu+1}}\varphi^{(3)}\right) \ ,
\label{eq:D.k.x.series}
\end{eqnarray}
which is valid for any  \emph{real} $\nu$, as we explain in
Appendix \ref{App:A.3}.
From the above equation, one can see that the defined operator
$D^{(\nu)}$ reproduces at small $x$ and/or large $\nu$
the  derivative of $\varphi(x)$ at the origin $x=0$.
Strictly speaking, using (\ref{eq:D.nu.x}) and (\ref{eq:D.k.x.seriesA}),
one obtains
$\lim\limits_{x\to 0}   [D^{(\nu+2)}\varphi](x)=\varphi'(0)$
(at fixed $\nu\in \mathbb{R}$)
and
$\lim\limits_{\nu\to \infty}   [D^{(\nu+2)}\varphi](x)=\varphi'(0)$
(at fixed $x$).
For this reason, we may appeal to Eq.\ (\ref{eq:int-dev-it}) and call
the variation range of the operator $D^{(\nu+2)}$ the
``integral derivative'' of $\varphi$.
Having defined this operator, we can derive the following expression
\begin{eqnarray}
  [D^{(\nu+2)}\varphi](x)
=
  \frac{1}{x} \int\limits_0^x\!\!\varphi(y) f(y,\nu,x)\,dy \ ,
  \label{eq:D.nu.x}
\end{eqnarray}
where
\begin{eqnarray}
  f(y,\nu,x)
=
  \frac{\theta(x-y)}{\Gamma(\nu+1)\, y}
  \left(\ln \frac{x}{y}\right)^\nu
\label{eq:f.n.x}
\end{eqnarray}
for any real $\nu $ (see Appendix \ref{App:A.3}).
As it is seen from Eq.\ (\ref{eq:D.nu.x}), the function $f(y,\nu,x)$
acts as a ``smooth projector'' onto the vicinity of the origin of
$y$, as one can appreciate from Fig.\ \ref{fig:fykx} in Appendix
\ref{App:A.3}.

By applying the operator $[D^{(\nu+2)}]$ on both sides of the QCD
SR given by (\ref{eq:pionDAQCDSR}), we obtain a new SR, viz.,
\begin{equation}
\begin{split}
&   f_{\pi}^2\,[D^{(\nu+2)}\varphi_\pi](x)
  + f_{A_1}^2\,{\rm e}^{-m_{A_1}^2/M^2}[D^{(\nu+2)}\varphi_{A_1}](x)
  + \int\limits_{s_0}^{\infty}
      [D^{(\nu+2)}\rho_\text{pert}]\left(x\right) {\rm e}^{-s/M^2}ds
  \\
& =
  \int\limits_{0}^{\infty}
  [D^{(\nu+2)}\rho_\text{pert}]\left(x\right) {\rm e}^{-s/M^2}ds
  + [D^{(\nu+2)}\Delta\Phi_\text{G}](x,M^2)
  + [D^{(\nu+2)}\Delta\Phi_\text{V}](x,M^2)  \\
& ~~~~~~~~~~~~~~~~~~~~~~~~~~~~~~~~~~~~~~~
  + [D^{(\nu+2)}\Delta\Phi_\text{T}](x,M^2)
  + [D^{(\nu+2)}\Delta\Phi_\text{S}](x,M^2)
  \, .
\end{split}
\label{eq:pionDAQCDSR:Dk}
\end{equation}
In order to achieve a better stability of this SR, we take into
account an effective $A_1$-meson state that embodies the $\pi'$
and the $a_1$ mesons and has the decay constant
$f_{A_1}=0.227$~GeV and the mass $m_{A_1}^2=1.616$~GeV$^2$.
For the pion-decay constant and the continuum threshold, we use
$f_\pi=0.137$~GeV and $s^\text{NLO}_0=2.25$~GeV$^2$, respectively.
These values were derived before from the corresponding two-point
QCD SRs with nonlocal condensates, see Ref.\ \cite{BMS01}.
However, because in our SR---cf.\ (\ref{eq:pionDAQCDSR:Dk})---we have
to resort to the LO expression for the spectral density (recall what
we said about this issue before), we adopt a somewhat larger value
of the threshold parameter:
$s_0\approx s^\text{NLO}_0(1+\alpha_S/\pi)=2.61$~GeV$^2$.
This is done for both the integral as well as the differential SRs,
the reason being that we want to preserve the correct normalization
of the pion DA.\footnote{We thank A.~P.~Bakulev for useful remarks
on this point.}

\begin{figure}[ht]
 \centerline{   \includegraphics[width=0.5\textwidth]{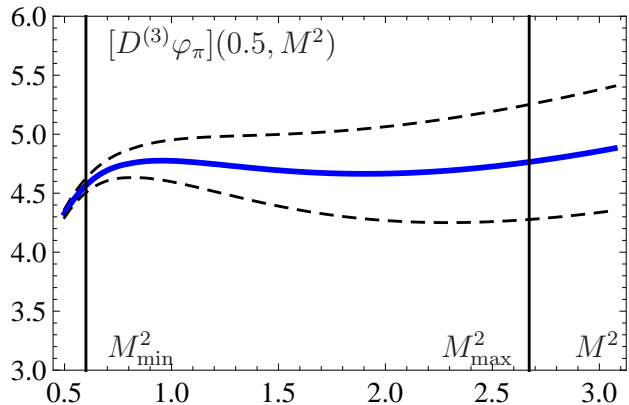}  
              }
  \caption{\footnotesize
  Dependence of $[D^{(3)}\varphi_\pi](x,M^2)$ on the auxiliary Borel
  parameter $M^2$ for $x=0.5$.
  The solid blue line corresponds to $[D^{(3)}\varphi_{A1}](x)=6.4$,
  whereas the dashed lines refer to $[D^{(3)}\varphi_{A1}](x)=6.7$
  (upper curve) and $[D^{(3)}\varphi_{A1}](x)=7.1$ (lower curve),
  using in all cases $x=0.5$.
  The two vertical lines delimit the fidelity region
  $M^2\in[M^2_\text{min},M^2_\text{max}]$.
\label{fig:D.k1.x05.M2}}
\end{figure}

As usual, we study the SRs in the fiducial interval of the Borel
parameter
$M^2\in[M^2_\text{min},M^2_\text{max}]$,
where both terms, the continuum contribution and the nonperturbative
one, each contributes about $1/3$ to the whole SR
(\ref{eq:pionDAQCDSR}).
This induces an uncertainty of the order of $(1/3)^2\to10\%$.
Moreover, the quantities to be calculated with the SR (the integral
derivatives) should not (crucially) depend on the Borel
parameter.
Therefore, we should take care that this dependence is minimized.
To achieve this goal, we attempt to minimize the root-mean-square
deviation by varying the
$[D^{(\nu+2)}\varphi_{A1}](x)$
contribution in the fiducial Borel interval.
On this account, we can average the $M^2$-dependence of the pion DA
contribution [first term in Eq.\ (\ref{eq:pionDAQCDSR:Dk}))] in
order to get a more reliable form of
the SR.
Therefore, we write
\begin{eqnarray}
\!\!\!\!\nonumber
  [D^{(\nu+2)}\varphi^\text{SR}_\pi](x)
&=&
  \langle[D^{(\nu+2)}\varphi_\pi](x,M^2)\rangle \\
&\equiv&
  \frac{1}{M^2_\text{max}-M^2_\text{min}}
  \int_{M^2_\text{min}}^{M^2_\text{max}}\!\!dM^2\,
  [D^{(\nu+2)}\varphi_\pi](x,M^2) \ .
\label{eq:mean-DA}
\end{eqnarray}
In Fig.\ \ref{fig:D.k1.x05.M2} we show the $M^2$-dependence of
$[D^{(3)}\varphi_\pi](x,M^2)$, obtained from SR
(\ref{eq:pionDAQCDSR:Dk}), for different values of
$[D^{(3)}\varphi_{A1}](x)$ and evaluating it for $x=0.5$.
The average value in the fiducial interval is
$[D^{(3)}\varphi_\pi^\text{SR}](0.5)=4.8 \pm 0.5$.

\begin{figure}[h!]
 \centerline{   \includegraphics[width=0.35\textwidth]{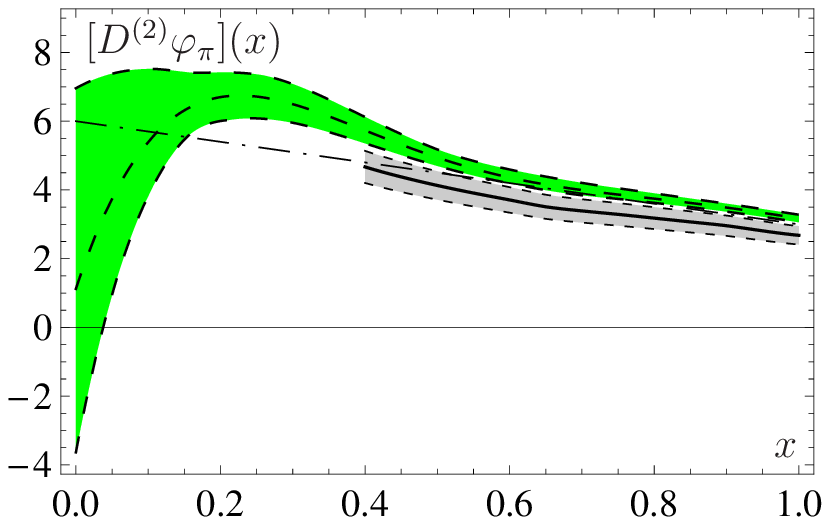}  
                \includegraphics[width=0.35\textwidth]{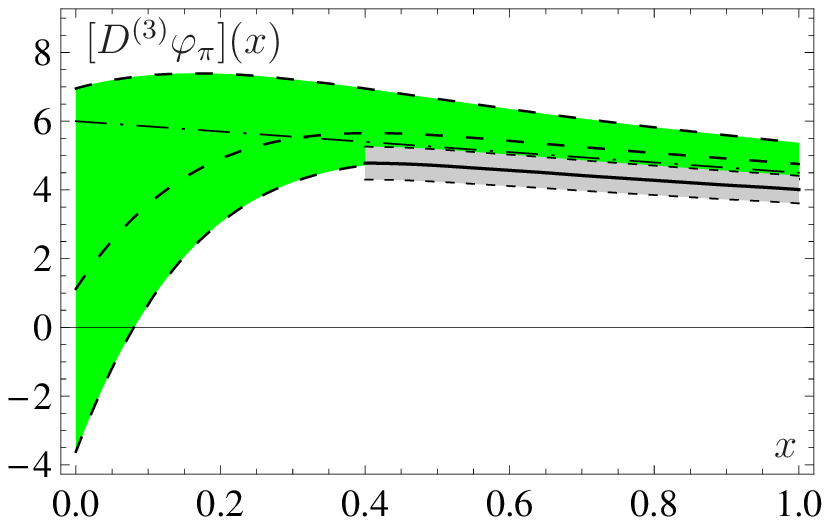}  
                \includegraphics[width=0.35\textwidth]{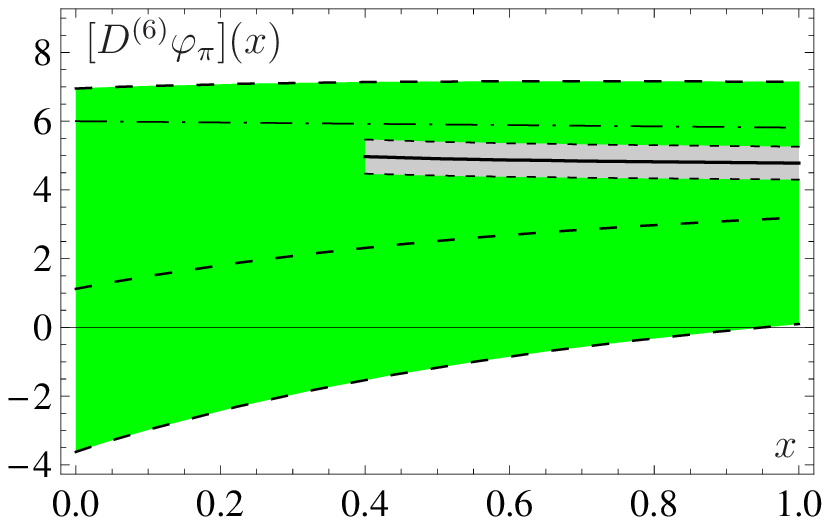}  
             }
  \caption{\footnotesize
  We show the $x$-dependence of $[D^{(\nu+2)}\varphi_\pi](x)$ for the
  BMS bunch of pion DAs \protect\cite{BMS01} (shaded green band within
  long-dashed lines) in comparison with the SR result
  (\ref{eq:mean-DA}) (narrow gray strip) in all three panels.
  The left panel shows the predictions for $\nu=0$, whereas those for
  $\nu=1$ and $\nu=4$ are shown in the middle and the right panel,
  respectively.
  The dashed-dotted line represents the asymptotic result
  $[D^{(\nu+2)}\varphi^\text{as}](x)=6-3x/2^\nu$.
\label{fig:DDDA012}}
\end{figure}

We discuss now the integral sum rules and their applications.
The main contribution from the singularities in the SR for
$[D^{(\nu+2)}\varphi_\pi](x,M^2)$ stems from the $x$-region around
$\Delta\equiv\lambda_q^2/(2 M^2)$.
This is because we used a delta-ansatz model [cf.\ (\ref{eq:ansv}),
(\ref{eq:anstril})] for the condensates, which implies that
the nonperturbative contributions have delta-function and
Heaviside-function behaved terms (\ref{eq:phi_s})--(\ref{eq:phi.G}).
Therefore, in order to take into account all NLC contributions,
we should analyze the region $x\gtrsim 0.4$ and use
$M^2_\text{min}\geq 0.6$~GeV$^2$
which corresponds to
$\Delta\leq 1/3< x$.
Moreover, the image of the operator $D^{(\nu+2)}$ for $\nu \geq 4$ is
numerically very close to the result obtained with the differentiation
method (see next subsection)---for any $x$.
Thus, the integral SR (\ref{eq:pionDAQCDSR:Dk}) becomes close to
the differential SR which we will consider in the next section.
For these reasons, we analyze the constructed SR
(\ref{eq:pionDAQCDSR:Dk}) for $\nu=0,1,2,3,4$ and $x>0.4$ and present
the results in Fig.\ \ref{fig:DDDA012} by the solid line that is inside
the light gray strip bounded by the short-dashed lines.
For the sake of comparison, the predictions for the asymptotic
DA (dashed-dotted line) and the BMS DA bunch---obtained in the NLC SR
analysis of Ref.\ \cite{BMS01}---(shaded band limited by long-dashed
lines) are also shown.
From this figure we see that our SR estimates for
$[D^{(\nu+2)}\varphi^\text{SR}_\pi](x)$ agree fairly well with the BMS
model---see also Table \ref{tab:all-derivatives}.
This table shows estimates for the third-order integral derivative of
the pion DA for $x=0.5$, using (i) the sum rule given by Eq.\
(\ref{eq:pionDAQCDSR:Dk}) and (ii) the pion DA models we discussed
above, and also flat-type DAs which we consider below.

We turn now our attention to flat-type DAs.
First, we compare the QCD SR result, obtained in (\ref{eq:mean-DA}),
with what one finds with the flat-top model
$\varphi^\text{flat}_\text{(B.1)}(x)\sim (x(1-x))^\alpha$
given in Appendix \ref{App:DAmodels}.
For this model, one has
\begin{eqnarray}\nonumber
  [D^{(\nu+2)}\varphi^\text{flat}_\text{(B.1)}](x)\gg
&
  [D^{(\nu+2)}\varphi^\text{as}](x)
\gtrsim
  [D^{(\nu+2)}\varphi_\pi^\text{SR}](x)
  ~~~~\text{for $\alpha < 0.1$}
\end{eqnarray}
for any real $\nu\in \mathbb{+R}$ and $0<x<1$
For the value $\alpha=0.1$, we find
$[D^{(3)}\varphi^\text{flat}_\text{(B.1)}](0.5)=227$, which is much
larger and far outside the range of values extracted from our SR.

Second, we consider a particular flat-type pion DA which is
provided by the AdS/QCD correspondence in the holographic
approach---see, for instance, Refs.\ \cite{BT07,GR08,KL08,AGN08}.
In that case one has $\alpha=0.5$ yielding
$[D^{(3)}\varphi^\text{hol}](0.5) =14$.

\begin{figure}[h!]
\centerline{\includegraphics[width=0.45\textwidth]{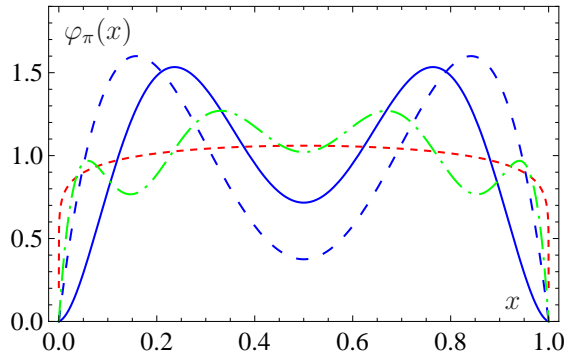}   
           }
\caption{\footnotesize
Comparison of selected pion DA models.
Solid blue line---central line of the BMS bunch \protect\cite{BMS01};
long-dashed blue line---CZ model\protect\cite{CZ84};
dashed-dotted green line---flat-like model given by (\ref{eq:3GegeDA}),
dotted red line---flat-top DA from Eq.\ (\ref{eq:flat-DA})
with $\alpha=0.1$.
All DAs are normalized at the same scale $\mu_0^2\simeq 1$~GeV$^2$
with or without evolution \cite{BMS01}.
\label{fig:pionDAmodels}}
\end{figure}

Third, we study an alternative flat-like pion DA which
results from the Gegenbauer expansion of unity by retaining only the
first few harmonics.
One obtains
\begin{eqnarray}
  \varphi^\text{flat}_\text{(\ref{eq:3GegeDA})}(x)
&=&
  6x\bar x
  \sum_{n=0}^{3}C^{3/2}_{2n}(2x-1)
  \frac{2(4n+3)}{3(2n+1) (2n+2)}
\label{eq:3GegeDA}
\end{eqnarray}
with a profile shown in Fig.\ \ref{fig:pionDAmodels} in comparison
with the models already mentioned:
BMS---solid line;
CZ---long-dashed blue line;
flat-top DA given by Eq.\ (\ref{eq:flat-DA})---dotted red line;
flat-like DA given by Eq.\ (\ref{eq:3GegeDA})---dashed-dotted green
line.

It is interesting to notice that using the pion DA given by expression
(\ref{eq:3GegeDA}) to calculate the pion-photon transition form factor
according to Radyushkin's expression (24) in Ref.\ \cite{Rad09}, one
actually reproduces the gross features of his results.
This proves that the inclusion of the first few Gegenbauer polynomials
in (\ref{eq:3GegeDA}) does not affect the result obtained with
$\varphi(x)=1$ in the momentum range $Q^2\leq 40$~GeV$^2$ in a crucial
way.
This type of DA, i.e., (\ref{eq:3GegeDA}), yields for the integral
derivative the value
$[D^{(3)}\varphi^\text{flat}_\text{(\ref{eq:3GegeDA})}](0.5)=22.5$,
which is much larger than the range of values determined via our SR.
This holds true also for the other two flat-type DAs considered above.
Recalling that the leading-order QCD sum rules with the minimal
Gaussian model for the nonlocal condensates provide much smaller
values of the integral derivative of the pion DA, one may conclude that
it is very difficult to reconcile flat-type pion DAs with SR
(\ref{eq:pionDAQCDSR:Dk}).

On the other hand, also the CZ model yields third-order integral
derivatives which are incompatible with the values derived from our SR
(\ref{eq:pionDAQCDSR:Dk})---see Table \ref{tab:all-derivatives}.
Note that a similar statement also applies to the pion DA
proposed in \cite{WH10}, which employs a Brodsky-Huang-Lepage
ansatz for the $\mathbf{k}_\perp$-dependence of the pion wave
function---see Table \ref{tab:all-derivatives}.
The upshot of this table is that the SR for the integral derivative
of the pion DA is fulfilled by the BMS bunch, whereas flat-type DAs
have no overlap with the estimated range of values.
The same is true for the CZ model.

\newcommand{\pha}{\strut$\vphantom{\vbox to 6mm{}}\vphantom{_{\vbox to 4mm{}}}$}         
\begin{table}[thb]\vspace*{-3mm}
\caption{
Results for the third-order integral derivative for $x=0.5$ and the
(usual) derivative of the pion DA, using different SR approaches
(first three rows) and pion DA models (six last rows).
\label{tab:all-derivatives}\vspace*{+1mm}}
\begin{ruledtabular}
\begin{tabular}{|c|l|c|c|} 
\pha  ~~~& Approach/Model                           & ``Integral'' derivative $[D^{(3)}\varphi_\pi](0.5)$& Derivative $\varphi'_\pi(0)$        \\ \hline
\pha   1 & Integral SR (\ref{eq:pionDAQCDSR:Dk})    & $4.7 \pm 0.5$                                             & $5.5 \pm 1.5$                \\ \hline
\pha   2 & Differential SR (\ref{eq:DpionDAQCDSR})  & ---                                                       & $5.3 \pm 0.5$                \\ \hline
\pha   3 & SR (\ref{eq:DpionDAQCDSR}) with
           smooth NLC (\ref{eq:smoothmodel})        & ---                                                       & $7.0 \pm 0.7$                \\ \hline\hline
\pha   4 & BMS bunch~\cite{BMS01}                   & $5.7\pm 1.0$                                              & $1.7 \pm 5.3$                \\ \hline
\pha   5 & Asymptotic DA                            & $5.25$                                                    & $6$                          \\ \hline
\pha   6 & CZ DA~\cite{CZ82}                        & $15.1$                                                    & $26.2$                       \\ \hline
\pha   7 &    DA from \cite{WH10}                   & $14$                                                      & $0$                          \\ \hline
\pha   8 & Flat-type DA, Eq.\ (\ref{eq:3GegeDA})    & $22.5$                                                    & $72$                         \\ \hline
\pha   9 & flat-top DA
           (Eq.\ (\ref{eq:flat-DA}), $\alpha=0.1$)  & $227$                                                     & $\gg 6$                     
\end{tabular}
\end{ruledtabular}
\end{table}

We close this subsection by considering the usual derivative
$\varphi_\pi'(0)$ of the pion DA which encapsulates the key
characteristics of the pion DA at small $x$.
This quantity can be extracted from Fig.\ \ref{fig:DDDA012}, where we
have plotted the results for $[D^{(\nu)}\varphi^\text{SR}_\pi](x)$
we obtained with the SR (\ref{eq:pionDAQCDSR:Dk}) for different
values of $\nu$.
To determine $\varphi_\pi'(0)$ one can use expansion
(\ref{eq:D.k.x.series})
\begin{eqnarray}
      [D^{(\nu+2)}\varphi^\text{SR}_\pi](x)
&\approx&
       \varphi'_\pi(0) + \varphi''_\pi(0)\frac{x}{2!2^{\nu+1}}
\label{eq:DDA00}
\end{eqnarray}
and subtract the second derivative for which the asymptotic value
$\varphi''_\pi(0)= -12(6)$ is used.
The involved error $\pm 6$ was estimated by varying (\ref{eq:DDA00})
within the narrow grey strip in Fig.\ \ref{fig:DDDA012}.
To be more specific, one obtains
\begin{eqnarray}
      \varphi'_\pi(0)
\approx
      [D^{(\nu+2)}\varphi_\pi^\text{SR}](x)
      - \varphi''_\pi(0)\frac{x}{2!2^{\nu+1}}
\approx
      [D^{(\nu+2)}\varphi_\pi^\text{SR}](x)
      +\frac{3x}{2^{\nu}}=5.5 \pm 1.5
\label{eq:DDA0}
\end{eqnarray}
for any $0.4<x$ and $0 \leq \nu \leq 4$.
The error in (\ref{eq:DDA0}) is a combination of the uncertainties
originating from SR (\ref{eq:pionDAQCDSR:Dk}) and the error in the
determination of $\varphi''_\pi(0)$.

The above finding can be compared with what one obtains for the BMS
 and the CZ model (displayed in Table \ref{tab:all-derivatives})
\begin{eqnarray}
   \varphi_\pi'(0)
&=&
   6\left[1+6\,a^{\text{BMS}}_2 + 15\,a^{\text{BMS}}_4\right]
=
  1.07^{+5.87}_{-4.68}=-3.61 \div 6.95
\label{eq:Der.DA.BMS}
\end{eqnarray}
and
\begin{eqnarray}
  \varphi_\pi'(0)
=
  6\left[1+6\,a^{\text{CZ}}_2\right]\simeq 26.2 \ ,
\label{eq:Der.DA.CZ}
\end{eqnarray}
respectively, using in both cases the normalization scale
$\mu^2\simeq 1$~GeV$^2$.
Quite analogously to the integral derivative, the CZ model
yields also for the standard derivative much larger values than those
estimated in (\ref{eq:DDA0}).
By contrast, the pion DA model proposed in \cite{WH10}---though it
provides a similarly large integral derivative like the CZ DA---has a
usual derivative at the origin which is zero due to the strong
exponential suppression of this DA in the vicinity of the origin.

\subsection{Differential sum rules}
\label{subsec:dif-SR}

Another way to study the behavior of the pion DA in the small-$x$
region is provided by the differentiation of the SR
(\ref{eq:pionDAQCDSR}), which yields
\begin{eqnarray}
  f_{\pi}^2\,\varphi_\pi'(0,M^2)
=
  \frac{3}{2\pi^2}M^2\left(1-{\rm e}^{-s_0/M^2}\right)
  + 18 A_S \Phi'
  -f_{A_1}^2\,\varphi_{A_1}'\!(0)\,{\rm e}^{-m_{A_1}^2/M^2} \ .
\label{eq:DpionDAQCDSR}
\end{eqnarray}
We shall evaluate this SR for the threshold value $s_0=2.61$~GeV$^2$,
recalling that we employing a LO expression for the spectral density.

Using the simplest delta-ansatz model for the condensates (cf.\
(\ref{eq:anstril})), only one nonperturbative term survives, namely,
the four-quark-condensate
\begin{eqnarray}
  \Phi'
=
  \frac{1}{18 A_S} \frac{d}{dx} \Delta\Phi_{S}(x,M^2)\Big |_{x=0}\ ,
\label{eq:4Qterm0}
\end{eqnarray}
where $\Delta\Phi_{S}(x,M^2)$ is represented by
Eq.\ (\ref{eq:Delta.Phi.S})---see Fig.\ \ref{fig:4Q} in Appendix
\ref{App:A.1}.
The vector-quark condensate (V) and the gluon condensate (G)---recall
Eq.\ (\ref{eq:pionDAQCDSR})---give zero contributions in the region
$x<\Delta$, where $\Delta=\lambda_q^2/(2M^2)>0$ with
$\lambda_q^2=0.4$~GeV$^2$.
On the other hand, the antiquark-gluon-quark condensate (T) amounts to
a vanishing contribution in the region
$x<\min\left\{\Delta,1-2\Delta\right\}$.
For $M^2>0.4$~GeV$^2$ (or equivalently $\Delta<1/2$) this contribution
can also be neglected.

Even if we assume a behavior of the various condensates differing from
the delta-ansatz model, using, for instance, a smooth model like
(\ref{eq:smoothmodel}) for the scalar quark condensate (which implies
a decay at large distances not slower than the exponential decay---see
further below), the (V), (G), and (T) terms give only a small non-zero
contribution in the small-$x$ region.
Therefore, these terms can be neglected in first approximation in both
mentioned models.
In this section, we shall study the differential SR
(\ref{eq:DpionDAQCDSR}) using the mentioned models for the scalar
condensate.

The calculation of the nonlocal version of the four-quark-condensate
contribution involves only the scalar quark condensate.
The four-quark contribution $\Delta\Phi_{\text{S}}(x,M^2)$ was obtained
in \cite{MR86}  and can be used with various types of vacuum-quark
distributions $f_S(\alpha)$.
\begin{figure}[h]
 \centerline{\includegraphics[width=0.40\textwidth]{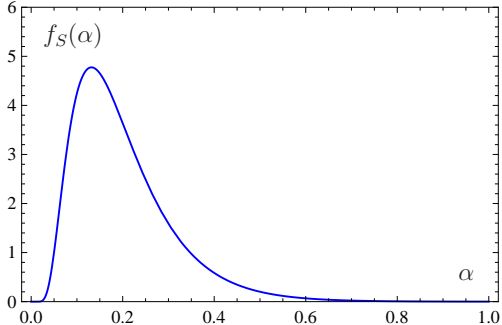} 
              }
  \caption{\footnotesize
          The distribution of vacuum quark virtualities in the smooth
          model (\ref{eq:smoothmodel}) for a particular choice
          of the intrinsic parameters as explained in the text.
   \label{fig:fS}}
\end{figure}
Accordingly, the probability for finding vacuum quarks with very large
virtualities is very small, as one can see from Fig.\ \ref{fig:fS}.
Therefore, the distribution function $f_S(\alpha)\simeq 0$ diminishes
in the region $\alpha>M^2$, the latter region corresponding to the
values $M^2>M^2_\text{min}\geq 0.6$~GeV$^2$ of the Borel parameter.
On the basis of this result, one \cite{BP06} can use the method of
Ref.\ \cite{MR86} to obtain the model-independent expression
(\ref{eq:Delta.Phi.S}).
Then, one finds for the four-quark-condensate contribution to the
SR (\ref{eq:DpionDAQCDSR}) the following expression
\begin{eqnarray}
   \Phi'
&=&
   \int_0^\infty\!\!\!d\alpha\frac{f_S(\alpha)}{\alpha^2}
=
   \langle{\bar{q}q}\rangle^{-1}\!\!\int_0^\infty\!\!
   z^2 M_S(z^2)dz^2 \ .
\label{eq:4Qterm}
\end{eqnarray}
We see from this equation that the nonperturbative contribution
to the SR is mainly due to the scalar-quark condensate
at large and moderate distances $z^2\sim 4/\langle k_q^2\rangle$.

For the concrete evaluation of the differential SR we again employ
the same criteria as already used in the integral SR for both the
continuum and the nonperturbative terms.
Applying these criteria, the low boundary for the Borel parameter
turns out to be very small, viz.,
$M^2_\text{min}=(0.3-0.4)$~GeV$^2$.
Though this low value is consistent with the applied criteria, it is
not in good agreement with standard QCD sum-rule approaches in which a
higher minimum Borel-parameter value is used.
Therefore, we use $M^2_\text{min}=0.6$~GeV$^2$,
a value also employed in the QCD SR for the pion DA in Ref.\
\cite{BMS01} in connection with the moments of the pion DA.
Keep also in mind that using a lower value of the Borel parameter
would cause the decrease of the first derivative of the pion DA at the
endpoints.
To continue, we define the derivative
$
 \varphi_\pi'(0)
=
 \langle \varphi_\pi'(0,M^2)\rangle
$
in terms of the mean value in the fiducial interval
$M^2\in[M^2_\text{min},M^2_\text{max}]$
following the definition on the RHS of (\ref{eq:mean-DA}).
This helps minimizing the sensitivity of $\varphi_\pi'(0) $ on the
choice of the Borel parameter by means of the variation of the $A_1$
contribution $\varphi_{A_1}'\!(0)$.

\begin{figure}[t]
 \centerline{\includegraphics[width=0.45\textwidth]{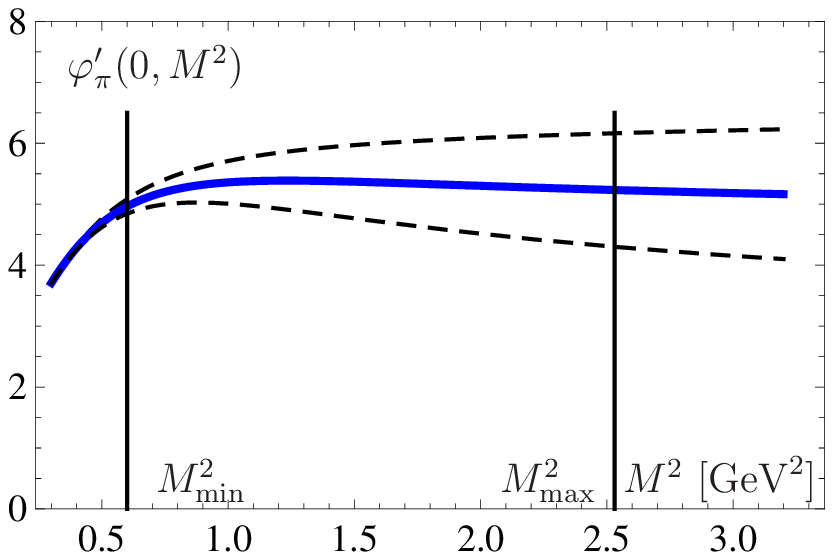}    
             ~~
             \includegraphics[width=0.45\textwidth]{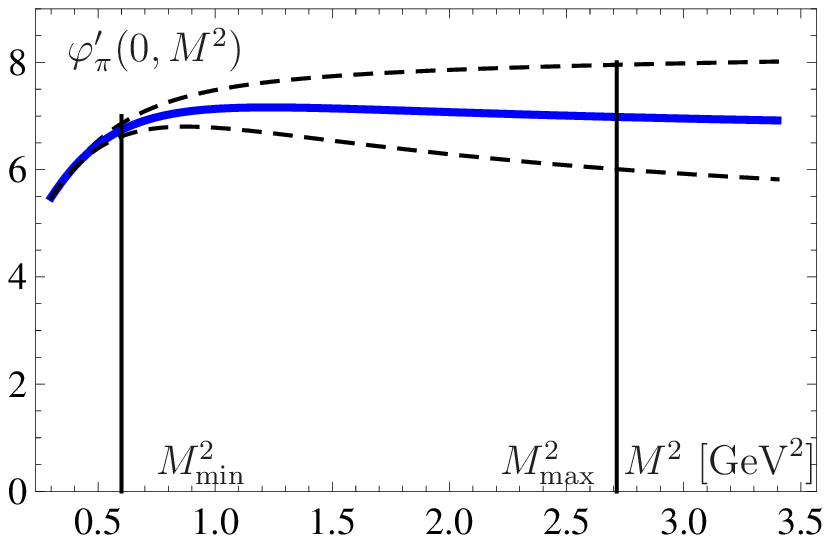}    
             }
  \caption{\footnotesize
    Left panel:
               The solid curve shows the $M^2$-dependence of $\varphi_\pi'(0,M^2)$
               in the SR with the $A_1$-meson contribution $\varphi_{A_1}'(0)=6.7$.
               The broken lines represent $\varphi_\pi'(0,M^2)$ for
               $\varphi_{A_1}'(0)=6.1$ (lower line) and $\varphi_{A_1}'(0)=7.5$ (upper line).
    Right panel:
                The solid curve illustrates the $M^2$-dependence of
                $\varphi_\pi'(0,M^2)$ in the SR which employs the smooth quark
                condensate model with a $A_1$-meson contribution given by
                $\varphi_{A_1}'(0)=6.8$.
                The broken lines denote
                $\varphi_\pi'(0,M^2)$ at $\varphi_{A_1}'(0)=6.1$ (lower line) and
                $\varphi_{A_1}'(0)=7.5$ (upper line).
\label{fig:SRM2}}
\end{figure}

The delta-ansatz
$
 f_S(\alpha)
=
 \delta(\alpha-\lambda_q^2/2)
$
leads to a simple expression for the nonperturbative contribution to
SR (\ref{eq:DpionDAQCDSR}), notably,
\begin{eqnarray}
  \Phi'
=
  \Phi_\text{delta}'
&=&
  \frac{4}{\lambda_q^4} \ .
\label{eq:delta-dev}
\end{eqnarray}
Note that we defined the $A_1$-meson contribution
$\varphi_{A_1}'(0)=6.7$ by means of the minimum of the root mean
square deviation.
On the other hand, the dependence of $\varphi_\pi'(0,M^2)$ on the
Borel parameter $M^2$ for the delta-ansatz model is controlled by
Eq.\ (\ref{eq:DpionDAQCDSR}) and is shown in the left panel of Fig.\
\ref{fig:SRM2}.
Thus, the average value of the pion DA derivative in the fiducial
Borel interval is $\varphi_\pi'(0)=5.3(5)$---see Table
\ref{tab:all-derivatives}.


We go forward and discuss the consequences of the smooth model for the
quark-virtuality distribution in the differential SR.

Though the delta-ansatz model is useful, because of its simplicity,
there is an indication from the heavy-quark effective theory
\cite{Rad91} that in reality the quark-virtuality distribution $f_S$
should be parameterized in a different way as to ensure that the
scalar condensate decreases exponentially at large distances.
Moreover, in order that the vacuum matrix element
$\langle{\bar{q}(D^2)^Nq}\rangle$ exists, the quark-virtuality
distribution $f_S(\alpha)$ should decrease faster than any power
$1/\alpha^{N+1}$ as $\alpha\to\infty$ \cite{MR86}.
For this reason, the authors of \cite{BM02,BM96} suggested a two-tier
model for $f_S$ which has a smooth dependence on the quark virtuality
$\alpha$, namely,
\begin{eqnarray}
  f_{S}(\alpha;\Lambda,n,\sigma)
&=&
  \frac{\left(\sigma/\Lambda\right)^{n}}{2K_n(2\Lambda\sigma)}\,
  \alpha^{n-1} e^{-\Lambda^2/\alpha-\alpha\,\sigma^2} \ ,
\label{eq:smoothmodel}
\end{eqnarray}
where $K_n(z)$ is the modified Bessel function.
This model, the so-called ``smooth model'', depends on two parameters
$\Lambda$ and $\sigma$
that parameterize, respectively, the long- and short-distance behavior
of the nonlocal condensates \cite{BM02}.
For large distances $|\,z|=\sqrt{-z^2}$ this model leads to the
asymptotic form
\begin{eqnarray}
  \langle{\bar{q}(0)q(z)}\rangle
&\stackrel{|\,z|\to\infty}{\longrightarrow}&
  \langle{\bar{q}q}\rangle|\,z|^{-(2n+1)/2}e^{-\Lambda|z|}
  \frac{2^{(2n-1)/2}\sqrt{\pi}\,\sigma^n}{\sqrt{\Lambda}\,
  K_n(2\Lambda\sigma)} \ .
\label{eq:smooth-x}
\end{eqnarray}

It is instructive to consider a purely exponential decay of the
quark-virtuality distribution and study its influence on the
quark condensate.
This can be realized in the model of \cite{BM02,BM96} by choosing $n=1$,
whereas the second parameter $\Lambda=0.45$ GeV can be taken from the QCD
SRs for the heavy-light meson in heavy quark effective theory---see
\cite{Rad91,Rad94}.
The two parameters $n$ and $\Lambda$ are responsible for the large-$z$
behavior of the scalar-quark condensate, cf.\ Eq.\ (\ref{eq:smooth-x}).
The third parameter $\sigma^2=10$ GeV$^{-2}$ is defined in terms
of the parameters $n, \Lambda$, and $\lambda_q^2$ via the following
equation:
\begin{eqnarray}
  \int_0^\infty\!\!\!\alpha\,
  f_{S}(\alpha;\Lambda,n,\sigma)\,d\alpha
=
  \frac{\Lambda}{\sigma}
  \frac{K_{n+1}(2\Lambda\sigma)}{K_n(2\Lambda\sigma)}
=
  \frac{\lambda_q^2}{2} \ ,
\label{eq:sigma-FS}
\end{eqnarray}
which we evaluate for the value of the nonlocality parameter
$\lambda_q^2=0.4$~GeV$^2$.
The main effect of using a smooth model for the quark-virtuality
distribution relative to the Gaussian form,
${f_S(\alpha)=\delta(\alpha-\lambda_q^2/2)}$,
is the induced increase of the nonperturbative contribution to the SR,
so that
\begin{eqnarray}
  \Phi_\text{smooth}'
&=&
  \int_0^\infty\!\!\!d\alpha
  \frac{f_{S}(\alpha;\Lambda,n,\sigma)}{\alpha^2}
=
  \frac{\sigma^2}{\Lambda^2}
  \frac{K_{n-2}(2\Lambda\sigma)}{K_{n}(2\Lambda\sigma)}
>
  \Phi_\text{delta}' \ .
\label{eq:smooth-1}
\end{eqnarray}

\begin{figure}[t]
 \centerline{\includegraphics[width=0.45\textwidth]{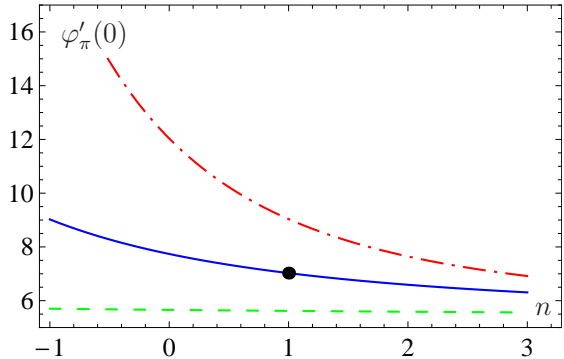}   
             }
  \caption{\footnotesize
   Dependence of the derivative $\varphi_{\pi}'(0)$ on the parameter
   $n$ of the smooth scalar condensate model (\ref{eq:smoothmodel})
   for $\Lambda=0.3 ~\text{GeV}$ (dashed-dotted red line),
       $\Lambda=0.45~\text{GeV}$ (solid blue line), and
       $\Lambda=1   ~\text{GeV}$ (dashed green line).
   Here, we show only the central value of the derivative
   $\varphi_{\pi}'(0)$ we obtained from the SR analysis
   (\ref{eq:DpionDAQCDSR}).
   The black dot symbol marks the position which corresponds to the model
   parameters $n=1$, $\Lambda=0.45~\text{GeV}$, and $\varphi_\pi'(0)=7.0(7)$,
   that corresponds to the third row in Table \ref{tab:all-derivatives}.
\label{fig:smoothSR.L.n}}
\end{figure}

We analyzed the SR (\ref{eq:DpionDAQCDSR}) for this model using a
particular choice of its parameters, notably,
$
 f_{S}(\alpha; \Lambda=0.45~\text{GeV}, n=1,
\sigma^2=10~\text{GeV}^{-2})
$
and determined the dependence of $\varphi_\pi'(0,M^2)$ on the Borel
parameter $M^2$.
The result is shown graphically in the right panel of Fig.\
\ref{fig:SRM2}.
The average value of the derivative $\varphi_\pi'(0,M^2)$ in the
fiducial Borel interval is $\varphi_\pi'(0)=7.0(7)$.
Thus, the nonperturbative contribution $\Phi_\text{smooth}'$,
obtained from the smooth model, is approximately two times larger
Xthan the analogous contribution $\Phi_\text{delta}'$ in the
delta ansatz:
$\Phi_\text{smooth}'\approx 2.3\,\Phi_\text{delta}'$.
In addition to this result, marked by a black dot, we show in Fig.\
\ref{fig:smoothSR.L.n} the dependence of $\varphi_\pi'(0)$
on the choice parameters $n$ and $\Lambda$ of the smooth model.
From this picture and the relation (\ref{eq:4Qterm}), we may come to
the conclusion that, choosing a model for the condensate that has a
slower decay at large distances (small $n$ or $\Lambda$), may cause an
increase of the nonperturbative contribution to the SR
(\ref{eq:pionDAQCDSR}) and entail also an increase of the value
$\varphi_\pi'(0)$.
On the other hand, choosing a model for the condensate with a faster
decay at large distances (large $n$ or $\Lambda$), may lead to a
decrease of the nonperturbative contribution to the SR
(\ref{eq:pionDAQCDSR}) and therefore to a decrease of the value
$\varphi_\pi'(0)$.

In the last column of Table \ref{tab:all-derivatives} we collect the
values of the (usual) pion DA derivative at $x\simeq 0$, using different
SR approaches (first three rows) and pion DA models (last six rows).

\section{Conclusions}
\label{sec:concl}

The current investigation was partly motivated by the recent
results obtained by the BaBar Collaboration \cite{BaBar09} on the
pion-photon transition form factor which indicate an unexpected growth
of this quantity with $Q^2$ above $\sim 10$~GeV$^2$ up to the highest
momentum value of 40~GeV$^2$ measured.
As it was pointed out in \cite{MS09} (see also
\cite{MS09Trento,MS09strba}), such behavior is impossible for
endpoint-suppressed pion DAs and brings into play a flat-type profile for
the pion DA, as proposed in Refs.\ \cite{Rad09,Pol09}.
Therefore, it appears to be of crucial importance to have a theoretical
tool in our hands able to reveal the particular characteristics of pion
DAs precisely in the kinematic endpoint region.

The method mostly used in the past to extract such information employs
the inverse
moment $\langle x^{-1}\rangle_\pi$---extensively discussed
in \cite{BMS01,BPSS04,BMS02,BMS03,BMS05lat}.
In the present work we proposed another, more direct way, to
access the endpoint characteristics of the pion DA which makes use of its
derivatives.
The first step was to define the notion of ``integral''
derivatives
$[D^{(\nu)}\varphi_\pi](x)$
by means of an appropriate operator $D^{(\nu)}$
[cf.\ (\ref{eq:D.nu.x})].
Next, we formulated an ``integral''  sum rule, Eq.\
(\ref{eq:pionDAQCDSR:Dk}), for these quantities by taking into
account an effective $A_1$-meson contribution.
Using this sum rule, we determined the range of values of the
integral derivatives and displayed it in the first entry
of the second column of Table \ref{tab:all-derivatives}
showing it also graphically in Fig.\ \ref{fig:DDDA012}.
We also calculated the integral derivative for different characteristic
pion DA models and listed its value in the same Table.
The list of pion DAs includes the BMS, the CZ, the asymptotic,
and two options for flat-type DA models, one given
by Eq.\ (\ref{eq:flat-DA}) with $\alpha=0.1$---``flat-top'' model---the
other being a flat-like model parameterized in terms of
Eq.\ (\ref{eq:3GegeDA}).
We also analyzed the usual derivative of the pion DA
in the vicinity of the origin.
This procedure is helpful in revealing the fine details of
the pion distribution amplitude in the endpoint region---see Table
\ref{tab:all-derivatives} for the results.

Our findings can be summarized as follows.
First, at the end of Subsec.\ \ref{subsec:int-SR},
we applied the result for the integral derivatives
$[D^{(\nu+2)}\varphi_\pi](x)$
in order to reproduce the usual derivative of $\varphi_\pi(x)$ at the
origin $x=0$.
The result $\varphi_\pi'(0)=5.5 \pm 1.5$ is shown in the third column
of Table \ref{tab:all-derivatives} in comparison with some
characteristic models for the pion DA.
Second, in Subsec.\ (\ref{subsec:dif-SR}), we studied the
derivative of the pion DA $\varphi_\pi'(0)$ in terms of the
differentiation of the SR (\ref{eq:pionDAQCDSR}) that leads to the
differential SR (\ref{eq:DpionDAQCDSR}).
It turns out that the only nonperturbative content in this SR is mainly
defined by the scalar quark condensate.
The nonperturbative contribution is proportional to the second inverse
moment (\ref{eq:4Qterm}) of the distribution $f_S(\alpha)$ of the
vacuum-quark virtualities and is defined by the behavior of the quark
condensate at large
and moderate distances between the vacuum quarks.
The results for the derivative of the pion DA $\varphi_\pi'(0)$ are
shown in Table \ref{tab:all-derivatives} for the delta-ansatz model
(\ref{eq:ansv}) and also for the two-tier smooth model
(\ref{eq:smoothmodel}) with the model parameters
$n=1$, $\Lambda=0.45~\text{GeV}$.
The dependence of the derivative of the pion DA $\varphi_\pi'(0)$ on
the choice of the model parameters $n$ and $\Lambda$ is illustrated in
Fig.\ \ref{fig:smoothSR.L.n}.

As we see from Table \ref{tab:all-derivatives}, the results from the
differential (\ref{eq:DpionDAQCDSR}) and the integral SRs
(\ref{eq:pionDAQCDSR:Dk}) agree with each other.
The integral derivative of the pion DA, based on a new SR
derived in this work, remains smaller than the asymptotic value and
overlaps with the range of values determined with the BMS bunch of pion
DAs, while there is no agreement with the CZ DA and the flat-type
models considered.
The same conclusions can be drawn also for the usual derivative of the
pion DA, which follows from the differential SR
(\ref{eq:DpionDAQCDSR}).
It is worth mentioning that employing the integral and the differential
sum rules, we found that the leading-order QCD sum rules
(\ref{eq:pionDAQCDSR}), which employ the minimal Gaussian model for the
nonlocal condensates, cannot be satisfied by flat-type pion
distribution amplitudes.

Given that an increasing behavior of the scaled pion-photon
transition form factor can only be achieved with flat-type pion DAs, the
independent experimental confirmation of this effect, e.g., by the BELLE
Collaboration, becomes extremely crucial for our theoretical
understanding of basic QCD exclusive processes.

\acknowledgments

We would like to thank Alexander Bakulev for stimulating discussions
and useful remarks.
A.P. is indebted to Prof. Maxim Polyakov for the warm hospitality
at Bochum University, where most of this work was carried out.
A.P. and S.M. acknowledge financial support from Nikolay Rybakov.
This work received partially support from the Heisenberg--Landau
Program under Grants 2009, and 2010, the Russian Foundation for
Fundamental Research (Grants No.\ 07-02-91557, No.\ 08-01-00686, and
No.\ 09-02-01149), and the BRFBR-JINR Cooperation Program, contract
No. F06D-002.
A.P. wishes to thank the Ministry of Education and Science of the Russian Federation
("Development of Scientific Potential in Higher Schools" projects: No.\ 2.2.1.1/1483 and No.\   2.1.1/1539),
and the DAAD Foundation for a research scholarship.

\begin{appendix}

\section{Expressions for the nonlocal contributions to the sum rules}
\renewcommand{\theequation}{\thesection.\arabic{equation}}
\label{App:A.1}

\setcounter{equation}{0}

In Sections \ref{sec:QCD-vac} and \ref{sec:pi-slope}, we used the
following expressions for the vacuum distribution functions
\begin{eqnarray}
  f_S(\nu)
&=&
  \delta\left(\nu-\lambda_q^2/2\right)\ ;\qquad
 f_V(\nu)\
=\
  \delta^\prime\left(\nu-\lambda_q^2/2\right)\ ;
\label{eq:ansv}\\
  f_{T_i}(\alpha_1,\alpha_2,\alpha_3)
&=&
  \delta\left(\alpha_1-\lambda_q^2/2\right)
  \delta\left(\alpha_2-\lambda_q^2/2\right)
  \delta\left(\alpha_3-\lambda_q^2/2\right)\ .
\label{eq:anstril}
\end{eqnarray}
The meaning of these expressions and their connection to the
initial NLCs has been discussed in detail in Refs.\ \cite{MR86,MR92}.
The contributions to the QCD SR (\ref{eq:pionDAQCDSR}) with nonlocal condensates,
$\Delta \Phi_{\Gamma}(x, M^2)$, associated with these expressions,
are shown below.
Here, and in what follows, we use
$\Delta \equiv \lambda_q^2/(2M^2)$,
$\bar\Delta\equiv 1-\Delta$.

Then we obtain
\newcommand{\xx}{\left(\bar x\rightarrow x\right)}
\begin{eqnarray}
  \Delta\Phi_S\left(x,M^2\right)
&=&
  \frac{A_S}{M^4}
  \frac{18}{\bar\Delta\Delta^2}
  \left\{
          \theta\left(\bar x>\Delta>x\right)
          \bar x\left[x+(\Delta-x)\ln\left(\bar x\right)\right]
         + \xx + \right. \nonumber \\
&&\qquad\qquad
  \left. + \theta(1>\Delta)\theta\left(\Delta>x>\bar\Delta\right)
         \left[\bar\Delta
              +\left(\Delta-2\bar xx\right)\ln(\Delta)\right]
  \right\} \ ,
\label{eq:phi_s}
\\
  \Delta\Phi_V\left(x, M^2\right) &=& \frac{A_S}{M^4}
    \left(x\delta'\left(\bar x-\Delta\right)+\xx\right)\ ,
\\
  \Delta\Phi_{T}\left(x,M^2\right)
&=&
  \Delta\Phi_{T_1}\left(x,M^2\right)
  +\Delta\Phi_{T_2}\left(x,M^2\right)
  +\Delta\Phi_{T_3}\left(x,M^2\right)\ ,
\nonumber
\end{eqnarray}
\begin{eqnarray}
  \Delta\Phi_{T_1}\left(x,M^2\right)
&=&
  -\frac{3 A_S}{M^4}
  \left\{
  \left[\delta(x-2\Delta) - \delta(x-\Delta)\right]
        \left(\frac1{\Delta} - 2\right)\theta(1>2\Delta)
        + \theta(2\Delta>x) \nn \right.\\ &&\left.
        \times \theta(x>\Delta)\theta(x>3\Delta-1)
        \frac{\bar x}{\bar\Delta}
        \left[\frac{3x}{\Delta} -6 - \frac{1+\bar x}{\bar\Delta}\right]
  \right\} + \xx \ ,
\label{eq:phi_t1}
\\
  \Delta\Phi_{T_2}\left(x,M^2\right)
&=&
  \frac{4 A_S}{M^4}\bar x
  \left\{\frac{\delta(x-2\Delta)}{\Delta}\theta(1>2\Delta)
           -\theta(2\Delta>x)\theta(x>\Delta)\theta(x>3\Delta-1)
           \right. \nonumber \\
&&\left.
          \times \frac{1+2x-4\Delta}{\bar\Delta\Delta^2}\right\}
  + \xx \ ,
\label{eq:phi_t2}
\\
  \Delta\Phi_{T_3}\left(x, M^2\right)
&=&
  \frac{3 A_S\bar x}{M^4\bar\Delta\Delta}
      \left\{\theta(2\Delta>x)\theta(x>\Delta)\theta(x>3\Delta-1)
      \left[2-\frac{\bar x}{\bar\Delta}-\frac{\Delta}{\bar\Delta}
      \right]\right\} \nn \\
& &
  + \xx\ ,
\\
  \Delta\Phi_G\left(x,M^2\right)
  &=&
  \frac{\langle \alpha_s GG \rangle}{24\pi M^2}
  \left(\delta\left(x-\Delta\right)+ \xx \right)\ .
\label{eq:phi.G}
\end{eqnarray}
In the above equations, we used the abbreviation
$
 \Ds A_S
=
 \frac{8\pi\alpha_S}{81}\langle\bar qq\rangle^2
$,
while for the quark and the gluon condensates the standard estimates
\cite{SVZ}
$
 \alpha_S\langle\bar qq\rangle^2
=
 1.83 \cdot10^{-4}$~GeV$^6
$,
$
 \Ds\frac{\langle\alpha_S GG\rangle}{12\pi}
=
 0.0012$ GeV$^4
$
and
$
 \Ds\lambda_q^2
=
 \frac{\langle\bar q\left(ig\sigma_{\mu\nu}G^{\mu\nu}\right)q\rangle}
 {2\langle\bar qq\rangle} =  0.4$~GeV$^2$,
normalized at $\mu^2 \approx 1$~GeV$^2$,
have been adopted.
In Sec.\ \ref{subsec:dif-SR}, we applied not only the delta ansatz,
discussed above, but also the smooth model, proposed in Refs.\
\cite{BM95,BM96}, which is based on an exponential decay of the
condensate (\ref{eq:smooth-x}).
For this reason, we use the model-independent
expression for the four-quark contribution \cite{MR86,BP06}
to obtain Eq.\ (\ref{eq:4Qterm})
\begin{eqnarray}\label{eq:Delta.Phi.S}
  \Delta\Phi_S\left(x,M^2\right)
&=&
  \frac{18A_S}{M^4}
  \mathop{\int\!\!\!\!\int}^{~~\infty}_{0\,0~}\!\!
  d\alpha_1\, d\alpha_2\,
  f_S(\alpha_1)\,
  f_S(\alpha_2)\,
\\\nonumber
&&
  \times \frac{x\,\theta \left(\Delta_1-\bar{x}\right)}
  {\Delta _1^2 \Delta _2 \bar{\Delta }_1^2}
  \left[\bar{x}\Delta_2\bar{\Delta}_1
  +\ln\left(\frac{x\Delta_1\bar{\Delta}_2}
  {x\Delta _1-(\Delta_1-\bar{x})\Delta_2}
  \right)
        \Delta_1(\Delta_1-\bar{x})\bar{\Delta}_2
  \right]
\\\nonumber
&&  + \xx \ ,
\end{eqnarray}
where $\Delta_i \equiv \alpha_i/M^2$,
$\bar\Delta_i \equiv 1-\Delta_i$, and $\bar x\equiv 1-x$.

\begin{figure}[h]
\centerline{\includegraphics[width=0.40\textwidth]{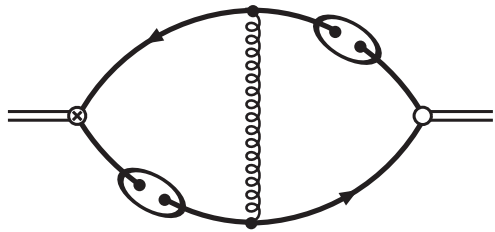}
             }
\caption{\footnotesize
        Typical diagram for the four-quark nonlocal-condensate $\Delta\Phi_\text{S}(x,M^2)$ encoding
        nonperturbative input in (\ref{eq:pionDAQCDSR}).
\label{fig:4Q}}
\end{figure}

\section{Modeling the pion DA}
\renewcommand{\theequation}{\thesection.\arabic{equation}}
\label{App:DAmodels}

In our work we compared the results following from the differential SR
(\ref{eq:DpionDAQCDSR}) for the usual derivative of the pion DA,
$\varphi_\pi'(0)$, with those obtained with the help of the integral derivative,
$[D^{(\nu)}\varphi_\pi](x)$, i.e., (\ref{eq:pionDAQCDSR:Dk}),
evaluating them for the BMS DA bunch, the CZ model, and a couple of
flat-type pion DAs.

The BMS bunch of pion DAs was obtained \cite{BMS01} as the sum of the
first three terms in the Gegenbauer-polynomial expansion
(\ref{eq:pion-DA}).
Their Gegenbauer coefficients were obtained from an analysis of the
SR (\ref{eq:pionDAQCDSR}) for the
$\langle\xi^{2N}\!\rangle$-moments ($N=0,1,2,3,4,5$)
with the nonlocal condensate contributions being presented in Appendix
\ref{App:A.1}.
It was found that only the two first Gegenbauer coefficients $a_2$
and $a_4$ give sizeable contributions to the
$\langle\xi^{2N}\!\rangle$-moments, whereas the higher Gegenbauer
terms give merely tiny contributions.
For this reason, the BMS bunch DAs are two-parameter models.
The central values of $a_2$ and $a_4$ for the whole BMS bunch in the
$a_2,a_4$ space are $a_2^{\rm BMS}= 0.187$ and $a_4^{\rm BMS}=-0.129$
using as a normalization scale {$\mu^2\simeq 1$~GeV$^2$}~\cite{BMS01}.

On the other hand, the CZ pion DA contains only the first nontrivial
Gegenbauer polynomial corresponding to the coefficient $a_2$
\cite{CZ84}.
It was derived from QCD SRs for the $\langle \xi^{2N} \rangle$-moments
($N=0,1,2$),
using local condensates.
For the sake of consistency, we use here a value of $a_2^{\rm CZ}=0.56$
obtained after evolution to the normalization scale
{$\mu^2\simeq 1$~GeV$^2$}, see for more details \cite{BMS02}.

One of the two flat-type models considered in this paper is
defined by
\begin{eqnarray}
  \varphi^\text{flat}_\text{(B.1)}(x)
&=&
   \frac{\Gamma(2(\alpha+1))}{\Gamma^2(\alpha+1)}x^\alpha(1-x)^\alpha
   \ .
\label{eq:flat-DA}
\end{eqnarray}
For this model, we find the following expressions:
\begin{eqnarray}
   [D^{(2)}\varphi^\text{flat}_\text{(B.1)}](x)
&=&
   \frac{\Gamma(2(\alpha+1))}{\Gamma^2(\alpha+1)}
   B_x(\alpha,1+\alpha)\ ,~~~~
   [D^{(\nu+2)}\varphi^\text{flat}_\text{(B.1)}](x)
\approx
   \alpha^{-\nu} [D^{(2)}\varphi^\text{flat}_\text{(B.1)}](x) \ .
\nn
\end{eqnarray}
These expressions can be generalized to any real differentiation index
$\nu\in \mathbb{+R}$ for $0<x<1$ to get
\begin{eqnarray}\nonumber
  [D^{(\nu+2)}\varphi^\text{flat}_\text{(B.1)}](x)&\gg
&
  [D^{(\nu+2)}\varphi^\text{as}](x)
\gtrsim
  [D^{(\nu+2)}\varphi_\pi^\text{SR}](x)
  ~~~~\text{for $\alpha \leq 0.1$}\ , \\ \nonumber
  227&\gg&  ~~~~~~~5.25~~~~~~ \gtrsim  4.7(5)
 \\\nonumber &&
  \text{for the particular values $\alpha=0.1$, $\nu=1$ and $x=0.5$}\ ,
\end{eqnarray}
\begin{eqnarray}\nonumber
  [D^{(\nu+2)}\varphi^\text{flat}_\text{(B.1)}](x)
&>&
  [D^{(\nu+2)}\varphi^\text{as}](x)
\gtrsim
  [D^{(\nu+2)}\varphi_\pi^\text{SR}](x)
  ~~~~\text{for $\alpha < 1$} \ , \\
  14&>&  ~~~~~~~5.25~~~~~~ \gtrsim  4.7(5)
 \\\nonumber &&
 \text{for the particular values $\alpha=0.5$, $\nu=1$ and $x=0.5$} \ .
\label{eq:D.k.flat2}
\end{eqnarray}

The other flat-type model, given by Eq.\ (\ref{eq:3GegeDA}),
was already discussed in the text.

\section{The spectral density}
\renewcommand{\theequation}{\thesection.\arabic{equation}}
\label{App:A.2}\setcounter{equation}{0}

The spectral density $\rho_\text{pert}\!\left(x\right)$ with NLO accuracy
was obtained in \cite{MR86,MR92} and was found to be
\begin{eqnarray}
  \rho_\text{pert}\!\left(x\right)
=
  3x\bar x
          \left[
                1 + \frac{\alpha_s}{4\pi}C_F
                \left(
                      5-\frac{\pi^2}{3}+\ln^2(\bar x/x)
                \right)
          \right]
          \frac{1}{2\pi^2} \ .
\label{eq:rho-NLO}
\end{eqnarray}
For our numerical calculations, which take into account radiative
corrections, we use the following function
\begin{figure}[b]
 \centerline{
             \includegraphics[width=0.45\textwidth]{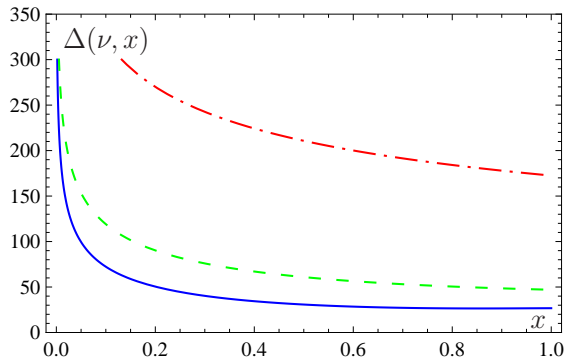}  
             }
             \vspace*{-5mm}
 \caption{\footnotesize
 $x$-dependence of the function $\Delta(\nu,x)$
 for $\nu=0$ (solid blue line), $\nu=1$ (dashed green line), and
 $\nu=4$ (dashed-dotted red line).
\label{fig:DeltaNLOvsLO}}
\end{figure}
\begin{eqnarray}
  \Delta(\nu,x)
&=&
  100 \frac{[D^{(\nu+2)}\rho_\text{pert}^\text{NLO}]\left(x\right)}
           {[D^{(\nu+2)}\rho_\text{pert}^\text{LO} ]\left(x\right)}
  -100 \ .
\label{eq:dnx}
\end{eqnarray}
The $x$-dependence of the function $\Delta(\nu,x)$ is illustrated
in Fig.\ \ref{fig:DeltaNLOvsLO} for different values of $\nu$:
solid blue line---$\nu=0$, dashed green line---$\nu=1$, and
dashed-dotted red line---$\nu=4$.

\section{Properties of the integral operator $D^{(\nu+2)}$}
\label{App:A.3}
\renewcommand{\theequation}{\thesection.\arabic{equation}}
\setcounter{equation}{0}

\begin{figure}[h!]
 \centerline{   \includegraphics[width=0.45\textwidth]{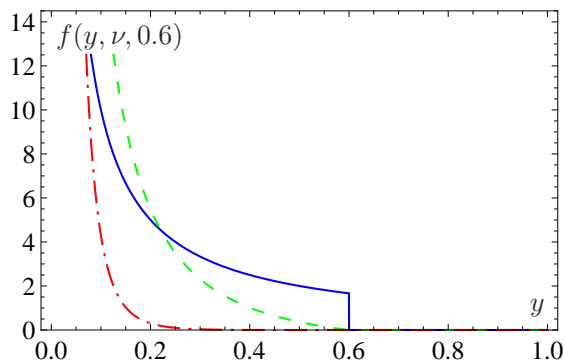}    
              }
  \caption{\footnotesize
            $y$-dependence of the function $f(y,\nu,0.6)$
            for $\nu=0$ (solid blue line), $\nu=1$ (dashed green line), and
            $\nu=4$ (dashed-dotted red line).
\label{fig:fykx}}
\end{figure}

In order to ensure a weak dependence of the results on the particular
model for the condensates adopted, and in order to include all
condensate contributions, one has to construct the SR by integrating the
pion DA SR (\ref{eq:pionDAQCDSR}) over a large enough interval of $x$.
For this reason, we introduced in Sec.\ \ref{subsec:int-SR} the
integral derivatives $[D^{(n)}\varphi](x)$, with the two lowest-order
ones being given in terms of Eq.\ (\ref{eq:int-dev012}).
The next higher derivative reads
\begin{eqnarray}
  [D^{(3)}\varphi](x)
=
  \frac{1}{x}\int\limits_0^x\frac{dy}{y}
  \int\limits_0^y\frac{\varphi(t)}{t}\,dt
=
  \frac{1}{x}\int\limits_0^x\frac{\varphi(t)}{t}dt
  \int\limits_t^x\frac{dy}{y}
=
  \frac{1}{x}\int\limits_0^x\frac{\varphi(t)}{t}
  \ln\left(\frac{x}{t}\right)dt \ .
\label{eq:D-3}
\end{eqnarray}
Assuming $\varphi(0)=0$, we find
\begin{eqnarray}
  [D^{(n+1)}\varphi](x)
=
  \frac{1}{x}\int\limits_0^x\!dy\,[D^{(n)}\varphi](x) \ .
\label{eq:D-n}
\end{eqnarray}
To obtain an expression for $[D^{(n+2)}\varphi](x)$ for any
$n\geq 0$ and $n\in \mathbb{N}$, one has to rearrange
the integration order to get
\begin{eqnarray}
  [D^{(n+2)}\varphi](x)
&=&
  \frac{1}{x}\int\limits_0^x\!\!\frac{dy_n}{y_n}
  \int\limits_0^{y_n}\!\!\frac{dy_{n-1}}{y_{n-1}}
\ldots
  \int\limits_0^{y_2}\!\!\frac{dy_{1}}{y_{1}}
  \int\limits_0^{y_1}\!\!\frac{\varphi(t)}{t}dt
=
  \frac{1}{x}\int\limits_0^{x}\!\!\frac{\varphi(t)}{t}
  \frac{1}{n!}\left(\!\int\limits_t^x\!\!\frac{dy}{y}\right)^n\!\!dt
  \ .
\label{eq:D-n-2}
\end{eqnarray}
This can be readily generalized to the expression (\ref{eq:D.nu.x})
using (\ref{eq:f.n.x}) that allows one to establish this transformation
for any real
$\nu$, $\nu\in \mathbb{R}$.
The dependence on $y$ of the function $f(y,\nu,x)$ is shown in
Fig.\ \ref{fig:fykx} for $x=0.6$ and $\nu=0,1,4$.
As one sees from Eq.\ (\ref{eq:D.nu.x}), the function
$f(y,\nu,x)$ acts as a smooth projector to the area around the origin
of $y$.

If a Taylor expansion exists for $\varphi(x)$ at $x=0$, then by
applying (\ref{eq:D.nu.x}) one finds
\begin{eqnarray}
  [D^{k+2}\varphi](x)
=
  \varphi'(0)+\varphi''(0)\frac{x}{2!2^{k+1}}
  + \sum\limits_{n=2}^\infty\varphi^{(n+1)}(0)
  \frac{x^n}{(n+1)!(n+1)^{k+1}} \ .
\label{eq:D.k.x.seriesA}
\end{eqnarray}
Using (\ref{eq:D.nu.x}) in combination with
(\ref{eq:D.k.x.seriesA}), one can obtain the
following properties of the operator $D^{(\nu)}$:
$
 \lim\limits_{x\to 0} [D^{(\nu)}\varphi](x)
=
 \lim\limits_{\nu\to \infty} [D^{(\nu)}\varphi](x)
=
 \varphi'(0)
$
and $D^{(\nu)}\left(a x+b x^2\right)=a + 2^{1-\nu}bx$.
Moreover, from (\ref{eq:D.k.x.seriesA}) it follows that the introduced
operator $[D^{(\nu)}\varphi](x)$ reproduces at small $x$ and large
$\nu$ the derivative of $\varphi(x)$ at $x=0$.

\end{appendix}


\end{document}